\documentclass[epj,refree,nopacs,draft]{svjour}
\def\D{\Delta}\def\T{\theta}\def\a{\alpha}\def\I{M_I}\def\U{M_U}
\def\Z{M_Z}\def\ov{\overline}\def\un{\underline}\def\op{\oplus}\def\t{\times}
\def\ep{\epsilon}\def\b{\begin{equation}}\def\e{\end{equation}}
\def\be{\begin{eqnarray}}\def\ee{\end{eqnarray}}\def\ba{\begin{array}}
\def\ea{\end{array}}
\begin{document}
\title{Precision and uncertainties in mass scale predictions in SUSY $SO(10)$ 
with $SU(2)_L\t SU(2)_R\t U(1)_{B-L}\t SU(3)_C$ intermediate breaking}
\author{M.K. Parida \thanks{\emph{e-mail:} mparida@sancharnet.in}
\and B. Purkayastha \and C.R. Das \thanks{\emph{e-mail:} crdas@email.com}
\and B.D. Cajee}\institute{Physics Department, North-Eastern Hill University, 
Shillong 793022, India}
\date{Received: date/ Revised version: date}
\abstract{In a class of  SUSY $SO(10)$ with $SU(2)_L\t SU(2)_R\t 
U(1)_{B-L}\t SU(3)_C$ $(g_{2L}\neq g_{2R})$ intermediate gauge symmetry, we 
observe that the prediction on the unification mass $(\U)$ is unaffected by 
Planck-scale-induced gravitational and intermediate-scale-threshold effects, 
although the intermediate scale $(\I)$ itself is subject to such corrections. 
In particular, without invoking the presence of additional lighter scalar 
degrees of freedom but including plausible and reasonable threshold effects, 
we find that interesting solutions for neutrino physics corresponding to 
$\I\simeq 10^{10}-10^{13}$ GeV and $\U\simeq (5-6)\t 10^{17}$ GeV are 
permitted in the minimal models. Possibilities of low-mass right-handed gauge 
bosons corresponding to $\I\simeq 1-10$ TeV consistent with the CERN-LEP data
are pointed out in a number of models when threshold effects are included 
using effective mass parameters.}
\maketitle
\section{Introduction}\label{sec1}
Supersymmetric grand unified theories (GUTs) have been the subject of 
considerable attention over the past two decades \cite{1,2,3,4}. While nonSUSY
$SU(5)$ fails to unify the gauge couplings of the standard model, 
$SU(2)_L\t U(1)_Y\t SU(3)_C (\equiv G_{213})$, the SUSY $SU(5)$ and single 
step breaking of almost all SUSY GUTs exhibit remarkable unification of gauge 
and Yukawa couplings at $\U\simeq 10^{16}$ GeV consistent with the recent 
CERN-LEP measurements. Compared to other GUTs, $SO(10)$ has several attractive
features. The fermions contained in the spinorial representation 
$\un{16}\subset SO(10)$ have just one extra member per generation which is
the right-handed neutrino needed to generate light Majorana neutrino masses
over a wide range of values through see-saw mechanism \cite{5}. It explains
why there is parity violation at low energies starting from parity
conservation at the GUT scale \cite{6,7}. It is the minimal left-right
symmetric GUT with natural quark lepton unification and having
$SU(2)_L\t SU(2)_R\t SU(4)_C$ \cite{7} as its maximal subgroup. It is has the 
potentiality to guarantee $R$-parity conservation in the Lagrangian.\par
With $\U\simeq M_N\simeq 10^{16}$ GeV, where $M_N=$ degenerate right-handed
Majorana neutrino mass, the grand desert model through see-saw mechanism
predicts much smaller values of light left-handed Majorana neutrino masses
than those needed for understanding neutrinos as hot dark matter (HDM)
candidate along with experimental indications on atmospheric neutrino
oscillations and neutrinoless double $\beta$ decay \cite{8,9}, unless
substantially lower values of $M_N$ are obtained by a judicious dialing of the 
Yukawa coupling of the right-handed Majorana neutrino, or via nonrenormalizable
operators. However, in such cases, one of the most attractive features like 
$b-\tau$ Yukawa unification for smaller values of $\tan\beta$ has to be 
sacrificed \cite{10,11}. On the other hand, SUSY $SO(10)$ with an intermediate 
gauge symmetry like $SU(2)_L\t SU(2)_R\t U(1)_{B-L}\t SU(3)_C (\equiv 
G_{2213})$ \cite{12,13,14,15,16} or $SU(2)_L\t SU(2)_R\t SU(4)_C (\equiv 
G_{224})$ \cite{17,18,19,20}, while providing a more natural value for $M_N$, 
substantially lower than the GUT scale, has the potentialities to account for
the $b-\tau$ Yukawa unification at the intermediate scale $\I\simeq10^9
-10^{13}$ GeV. In this context it has been demonstrated that desirable values 
of $G_{2213}$-breaking scale with $\I\simeq10^9-10^{13}$ are possible provided 
that a number of scalar components of full $SO(10)$ Higgs representations are 
light with masses near the intermediate scale \cite{13,14}.\par
Gravitational corrections due to higher dimensional operators \cite{21,23} and 
threshold effects due to superheavy particles have been shown to influence the
GUT predictions significantly \cite{17,23,24,25}. Since neither the superheavy
masses contributing to threshold effects near the GUT scale, nor the
coefficients of the higher dimensional operators contributing to gravitational
corrections are determined by the grand unified theories, these corrections
add to the uncertainties and inaccuracies of the model predictions. In order
to remove such limitations of the GUTs, it is important to search for gauge 
symmetries and possible representations for which some of the uncertainties
could be absent. For example, it has been demonstrated through theorems that 
in all GUTs with $SU(2)_L\t SU(2)_R\t SU(4)_C\t P (\equiv G_{224P}, g_{2L}=
g_{2R})$ intermediate symmetry, the GUT-threshold and all gravitational 
corrections on $\sin^2\T_W(\Z)$ and intermediate scale are absent 
\cite{18,19}. The presence of $G_{224P}$ intermediate gauge symmetry has been 
found to be essential for these cancellations.\par
The purpose of this paper is two fold. For the first time, we demonstrate 
certain precise results in a class of GUTs with $G_{2213}(g_{2L}\neq g_{2R})$ 
intermediate gauge symmetry with $D$-parity broken at the GUT scale \cite{26}.
In particular we find that in $SO(10)$ the dominant effect due to the 5-dim. 
operator is absent on $\U$ leading to the absence of such gravitational 
corrections on the proton lifetime for $p\to e^+\pi^0$. The threshold effects
caused due to the spreading of masses around the intermediate scale are also
found to be absent on $\U$. Secondly, while exploring uncertainties in the
intermediate scale predictions in SUSY $SO(10)$, we show, for the first time,
that the $G_{2213}$ intermediate symmetry is allowed to survive down to 
$\I\simeq 10^{10}-10^{13}$ GeV by threshold and gravitational corrections.
We have investigated the impact of threshold effect in SUSY $SO(10)$ models
with one pair of $\un{126}\op\un{\ov{126}}$ and one or two pairs of 
$\un{16}\op\un{\ov{16}}$ and find that even $\I\simeq 10$ TeV is allowed
consistent with the CERN-LEP measurements \cite{27} provided that the effective
mass parameters at the intermediate or GUT-thresholds are few times heavier
or lighter than the corresponding scales.\par
The paper is organized in the following manner. In Sec. \ref{sec2} we 
discuss the analytic formulas for mass scales including threshold and 
gravitational corrections. In Sec. \ref{sec3} we derive vanishing 
corrections due to 5-dim. operator on the GUT-scale and estimate gravitational
corrections on the intermediate scale. In Sec. \ref{sec4} we discuss the 
threshold effects and their impact on $\U$ and $\I$. The results are 
summarized with conclusions in Sec. \ref{sec5}.
\section{Analytic Formulas for Mass Scales}\label{sec2}
We consider the following symmetry breaking pattern and derive the analytic 
formulas for the unification mass $\U$ and the intermediate scale $\I$
including one-loop, two-loop, gravitational and threshold corrections.
\be &&SO(10)\t {\rm SUSY}\mathop{\longrightarrow}^{\un{210}}_{\U}G_{2213}\t 
{\rm SUSY}\nonumber\\ &&\mathop{\longrightarrow}^{S}_{\I}G_{213}\t {\rm SUSY}
\mathop{\longrightarrow}^{\un{10}}_{\Z}U(1)_{em}\t SU(3)_C\nonumber\ee
where the multiplet S is a component of the $SO(10)$ representations 
$\un{16}\op\un{\ov{16}}$ or $\un{126}\op\un{\ov{126}}$ as the case may be. 
The renormalization group equations in the presence of the two gauge symmetries
$G_{213}$ and $G_{2213}$ below the GUT scale can be written as
\be{1\over\a_i(\Z)}&=&{1\over\a_i(\I)}+{a_i\over 2\pi}\ln{\I\over\Z}+\T_i-\D_i,\nonumber\\ 
i&=&1Y, 2L, 3C;\label{eq1}\\ 
{1\over\a_i(\I)}&=&{1\over\a_i(\U)}+{a'_i\over 2\pi}\ln{\U\over\I}+\T'_i-\D'_i-\D_i^{NRO},
\nonumber\\ i&=&2L, 2R, BL, 3C.\label{eq2}\ee 
where the second term in the R.H.S. of eqs.(\ref{eq1}) and (\ref{eq2}) 
represents one-loop contributions and the third term of both the equations are
the two-loop terms \cite{28},
\be\T_i&=&{1\over 4\pi}\sum_jB_{ij}\ln{\a_j(\I)\over\a_j(\Z)},\nonumber\\
\T'_i&=&{1\over 4\pi}\sum_jB'_{ij}\ln{\a_j(\U)\over\a_j(\I)},\label{eq3}\\
B_{ij}&=&{b_{ij}\over a_j},\, B'_{ij}={b'_{ij}\over a'_j}.\label{eq4}\ee
While the functions $\D_i$ include threshold effects at $\Z$ and $\I$, 
$$\D_i=\D^{(Z)}_i+\D^{(I)}_i$$ $\D'_i$ include threshold effects at $\U$. 
The expressions for $\D_i$ and $\D'_i$ are given in Sec. \ref{sec5}. The term
$\D^{NRO}_i$ in eq. (\ref{eq2}) contains higher-dimensional-operator effects 
which modify the boundary condition at $\mu=\U$ as,\\
\be&&\a_{2L}(\U)(1+\ep_{2L})=\a_{2R}(\U)(1+\ep_{2R})\nonumber\\
&&=\a_{BL}(\U)(1+\ep_{BL})=\a_{3C}(\U)(1+\ep_{3C})=\a_G\label{eq5}\ee 
leading to
\b\D_i^{NRO}=-{\ep_i\over\a_G},\, i=2L, 2R, BL, 3C\label{eq6}\e
where $\a_G$ is the GUT-fine-structure constant. Considering the boundary 
condition (\ref{eq5}) along with eqs.(\ref{eq1}), (\ref{eq2}) and (\ref{eq6}) 
we obtain the following analytic formulas for mass scales.
\be\ln{\I\over\Z}&=&{1\over(AB'-A'B)}[(AL_S-A'L_\T)+(A'J_2-AK_2)\nonumber\\
&&-{2\pi\over\a_G}(A\ep''-A'\ep')+(A'J_\D-AK_\D)],\label{eq7}\\
\ln{\U\over\Z}&=&{1\over(AB'-A'B)}[(B'L_\T-BL_S)+(BK_2-B'J_2)\nonumber\\
&&-{2\pi\over\a_G}(B'\ep'-B\ep'')+(BK_\D-B'J_\D)],\label{eq8}\ee where
\be L_S&=&{2\pi\over\a(\Z)}\left(1-{8\over3}{\a(\Z)\over\a_S(\Z)}\right),\nonumber\\
L_\T&=&{2\pi\over\a(\Z)}\left(1-{8\over3}\sin^2\T_W(\Z)\right),\nonumber\\
A&=&a'_{2R}+{2\over3}a'_{BL}-{5\over3}a'_{2L},\nonumber\\
B&=&{5\over3}(a_Y-a_{2L})-\left(a'_{2R}+{2\over3}a'_{BL}-{5\over3}a'_{2L}\right),
\nonumber\\
A'&=&\left(a'_{2R}+{2\over3}a'_{BL}+a'_{2L}-{8\over3}a'_{3C}\right),\nonumber\\
B'&=&{5\over3}a_Y+a_{2L}-{8\over3}a_{3C}\nonumber\\
&&-\left(a'_{2R}+{2\over3}a'_{BL}+a'_{2L}-{8\over3}a'_{3C}\right),\nonumber\\
J_2&=&2\pi\left[\T'_{2R}+{2\over3}\T'_{BL}-{5\over3}\T'_{2L}
+{5\over3}(\T_{Y}-\T_{2L})\right],\nonumber\\
K_2&=&2\pi\left[\T'_{2R}+{2\over3}\T'_{BL}+\T'_{2L}-{8\over3}\T'_{3C}\right.
\nonumber\\
&&+\left.{5\over3}\T_{Y}+\T_{2L}-{8\over3}\T_{3C}\right],\nonumber\\
\ep'&=&\ep_{2R}+{2\over3}\ep_{BL}-{5\over3}\ep_{2L},\nonumber\\
\ep''&=&\ep_{2L}+\ep_{2R}+{2\over3}\ep_{BL}-{8\over3}\ep_{3C},\nonumber\\
J_\D&=&-2\pi\left[\D'_{2R}+{2\over3}\D'_{BL}-{5\over3}\D'_{2L}
+{5\over3}(\D_{Y}-\D_{2L})\right],\nonumber\\
K_\D&=&-2\pi\left[\D'_{2R}+{2\over3}\D'_{BL}+\D'_{2L}-{8\over3}\D'_{3C}\right.
\nonumber\\
&&\left.+{5\over3}\D_{Y}+\D_{2L}-{8\over3}\D_{3C}\right].\label{eq9}\ee
In the R.H.S. of eqs. (\ref{eq7}) and (\ref{eq8}) the first, second, third and
the fourth terms are one-loop, two-loop, gravitational and threshold 
contributions, respectively. The one-loop and the two-loop beta-function 
coefficients below the intermediate scale $(\I)$ are given by \cite{22,28},
\be&&\left(\ba{c} a_{Y}\\ a_{2L}\\ a_{3C}\ea\right)=\left(\ba{c} {33\over 5}\\ 
1\\ -3\ea\right),\nonumber\\
&&b_{ij}=\left(\ba{ccc} {199\over 25}&{27\over 5}&{88\over 5}\\ {9\over 5}&25&24\\ 
{11\over 5}&9&14\ea\right),\, i,j=1Y, 2L, 3C.\label{eq10}\ee
Above the intermediate scale, the one-loop and two-loop beta-function 
coefficients are
$$\left(\ba{c} a'_{2L}\\ a'_{2R}\\ a'_{BL}\\ a'_{3C}\ea\right)
=\left(\ba{c} n_{10}\\ n_{10}+n_{16}+4n_{126}\\ 6+{3\over 2}n_{16}+9n_{126}\\ 
-3\ea\right),$$ \be b'_{ij}&=&\left(\ba{cc} 18+7n_{10}&3n_{10}\\ 
3n_{10}&18+7n_{10}+7n_{16}+48n_{126}\\ 9&9+{9\over 2}n_{16}+72n_{126}\\ 
9&9\ea\right.\nonumber\\ &&\left.\ba{cc} 3&24\\ 3+{3\over 2}n_{16}+24n_{126}&24\\
7+{9\over 4}n_{16}+54n_{126}&8\\ 1&14\ea\right),\nonumber\\ 
i, j&=&2L, 2R, BL, 3C.\label{eq11}\ee
Including one- and two-loop corrections, we consider a variety of models taking
the lighter multiplet to be $$S=pn_{126}+qn_{16}$$
where p and q are integers. Here $n_{126}=1$ or $n_{16}=1$ imply that the 
components $\D_R(1,3,-1,1)+\ov\D_R(1,3,1,1)\subset \un{126}\op\un{\ov{126}}$ 
or $\chi_R(1,2,1/2,1)+\ov\chi_R(1,2,-1/2,1)\subset \un{16}\op\un{\ov{16}}$
of $SO(10)$ have masses close to $\I$. Here a minimal model is defined as the 
one with $n_{126}=1$ or $n_{16}=1$ where only one set of 
$\un{126}\op\ov{\un{126}}$ or $\un{16}\op\ov{\un{16}}$ is used for $G_{2213}$
breaking. In addition the GUT scale symmetry breaking is carried out by only
one representation like $\un{210}$ or $\un{45}$ which are needed for 
decoupling the parity and $SU(2)_R$-breakings. There are nonminimal models in
the literature as in refs. \cite{13,14} and in ref \cite{12}, the latter having
$n_{16}=3$. It may be noted that the spontaneous breaking of 
$SU(2)_R\t U(1)_{B-L}$ gauge symmetry by $\un{126}$ guarantees automatic
conservation of $R$-parity whereas the use of $\un{16}$ instead of $\un{126}$
leads to $R$-parity violation. In the latter case it is necessary to impose 
additional discrete symmetries to maintain the stability of the proton.
We use the following input parameters for our analysis \cite{27}
$$\a^{-1}(\Z)=128.9\pm0.09,\, \a_{3C}=0.119\pm 0.004,$$
\b\sin^2\T(\Z)=0.23152\pm 0.00032,\, \Z=91.187\, {\rm GeV}.\label{eq12}\e
Our solutions including only one-loop and two-loop contributions in different 
models are shown in Table \ref{tab1}. For example, if $n_{16}=1$ and 
$n_{126}=0$ the two-loop values are $\I=10^{16.9}$ GeV and $\U=10^{16.1}$ GeV. 
It is clear from Table \ref{tab1} that upto two-loop level the models do not 
allow $\I= 10^{15}$ GeV and in some cases $\I$ is even greater than $\U$ which
are forbidden. Also it is to be noted that $\U$ for all models attains a 
constant value of $10^{16.5}$ GeV. This phenomenon with occurrence of 
$\I\simeq\U$ has led to invoke the existence of lighter scalar degrees of 
freedom in order to bring down the value of the intermediate scale with 
$\I\ll\U$ \cite{13,14}.
\section{Gravitational Corrections on the Mass Scales}\label{sec3}
The mechanism of decoupling of parity and $SU(2)_R$-breakings is implemented in
$SO(10)$ by using the Higgs representation $\un{210}$ or $\un{45}$ for the
symmetry breaking at the GUT scale. Out of these two, the representation 
$\un{45}$ does not contribute to the gravitational corrections through the
5-dim. operator since $Tr(F_{\mu\nu}\Phi_{(45)}F^{\mu\nu})$ vanishes 
identically. Thus confining to the minimal model and using $\un{210}$ for the
$SO(10)$ symmetry breaking at the GUT scale, we demonstrate in this section
how the prediction on the unification mass has vanishing correction due to the
5-dim. operator. We also show how the gravitational effect lowers the 
intermediate scale by at most two orders of magnitude from the SUSY GUT-scale.
\subsection{Vanishing Gravitational Corrections on the Unification Scale}
\label{secc}
The super-Higgs representation $\un{210}$ contains the singlet $\xi(1,1,1)$ 
under $SU(2)_L\t SU(2)_R\t SU(4)_C$ which has been noted to be odd under 
$D$-symmetry that acts like the left-right discrete symmetry $(\equiv Parity)$ 
\cite{26}. But the neutral component in $\chi(1,1,15)$ of $\un{210}$ is even 
under the same $D$-symmetry. $SO(10)$ can be broken to $G_{2213}$ without 
left-right discrete symmetry by assigning vacuum expectation value 
$\langle\xi(1,1,1)\rangle=\langle\chi^0 (1,1,15)\rangle\simeq\U$. In this 
case it has been shown in ref. \cite{25} that the nonrenormalizable Lagrangian
containing the 5-dim. operator
\b-{\eta\over 2M_{Pl}}Tr\left(F_{\mu\nu}\Phi_{210}F^{\mu\nu}\right)\label{eq13}\e
yields via eqs.(\ref{eq5}) and (\ref{eq9}),
\be\ep_{2R}&=&-\ep_{2L}=-\ep_{3C}={1\over2}\ep_{BL}=\ep,\nonumber\\
\ep&=&{\eta\over 16}{\U\over M_{Pl}}\left[{3\over 2\pi\a_G}\right]^{1\over 2},\nonumber\\
\ep''&=&\ep_{2L}+\ep_{2R}+{2\over 3}\ep_{BL}-{8\over 3}\ep_{3C}=4\ep,\nonumber\\
\ep'&=&\ep_{2R}+{2\over3}\ep_{BL}-{5\over 3}\ep_{2L}=4\ep,\label{eq14}\ee
where $\a_G={1\over{24.3}}$. It is important to note that $\ep'=\ep''$
identically which has strong bearing on the prediction of the GUT scale. 
From eqs.(\ref{eq7}) and (\ref{eq8}) we have the gravitational corrections 
due to the 5-dim. operator,
\be\left(\ln{\I\over \Z}\right)_{NRO}&=&{2\pi(A'\ep'-A\ep'')\over\a_G(AB'-A'B)},
\label{eq15}\\
\left(\ln{\U\over \Z}\right)_{NRO}&=&{2\pi(B\ep''-B'\ep')\over\a_G(AB'-A'B)}.
\label{eq16}\ee
Now we demonstrate vanishing gravitational corrections to the unification mass
in the following manner. In all models with decoupled parity and $SU(2)_R$  
breakings where there are no additional $SU(2)_L$ or $SU(3)_C$-multiplets below
GUT scale, except the SM-Higgs doublets near $\Z$ and $\D_R\op\ov\D_R$ or
$\chi_R\op\ov\chi_R$ near $\I$ as the case may be,
\be a_{2L}&=&a'_{2L},\nonumber\\
a_{3C}&=&a'_{3C}.\label{eq17}\ee
Using eq. (\ref{eq17}) in eq.(\ref{eq9}), we obtain
\be B&=&B'={5\over 3}a_Y-{2\over 3}a'_{BL}-a'_{2R},\nonumber\\
A&=&a'_{2R}+{2\over 3}a'_{BL}-{5\over 3}a'_{2L},\nonumber\\
A'&=&a'_{2R}+{2\over 3}a'_{BL}+a'_{2L}-{8\over 3}a'_{3C}.\label{eq18}\ee
Eqs. (\ref{eq15}) and (\ref{eq16}) then yield with the help of (\ref{eq14}) and
(\ref{eq18}),
\be\left(\ln{\I\over \Z}\right)&=&{2\pi(A'\ep'-A\ep'')\over\a_GB(A-A')}
=-{8\pi\ep\over B\a_G},\nonumber\\
\left(\ln{\U\over \Z}\right)&=&{2\pi(\ep''-\ep')\over\a_G(A-A')}=0.\label{eq19}\ee
The result given in (\ref{eq19}) are valid both in SUSY and nonSUSY GUTs like
$SO(10), SO(18)$, and $E_6$ etc as long as $\ep'=\ep''$ as in (\ref{eq14}). 
This suggests an important aspect of the model that in $SO(10)$ with 
$D$-parity broken at the GUT scale, the GUT scale and the proton lifetime are 
unaffected due to gravitational corrections through the 5-dim. operator.
\subsection{Gravitational Correction on the Intermediate Scale}\label{secd}
We have examined the effect of gravitational correction on the intermediate
scale originating from the 5-dim. operator given in eq. (\ref{eq13}) in 
different models characterized by $(n_{16}, n_{126})=$ (0,1), (1,0), (1,1),
(2,0), (0,2), (2,1), (1,2), (0,3) and (3,0). The one-loop coefficient $A$, 
$A'$, $B$, $B'$ and numerical results obtained in different cases are shown in 
Table. \ref{tab1}. By varying $\eta$ parameter within the range $-10$ to 
$+10$, we obtain intermediate scale $\I$ between $10^{14}$ to $10^{16}$ GeV.
The maximal effects on $\I$ is found to occur in the minimal model with
$\un{210}$, $\un{126}\op\ov{126}$ and $\un{10}$ representations where
$R$-parity is automatically conserved and we obtain the lowest possible value
to be $\I\simeq10^{14}$ GeV.
\section{Threshold Effects}\label{sec4}
So far we have noted that the impact of gravitational corrections on 
intermediate scale $\I$ could bring it down to $10^{14}$ GeV whereas the 
unification scale remains unaffected. The possibility of 
$\I\simeq 10^{10}-10^{12}$ GeV with $b-\tau$ Yukawa unification at $\I$ has 
been addressed in ref \cite{13,14} but with a number of additional Higgs 
scalars having masses near $\I$, even though they do not contribute to the 
spontaneous symmetry breaking. But we demonstrate here that when threshold 
effects are taken into account, the scale $\I$ fits into the desired range of 
values even if gravitational corrections are ignored and there are no 
additional scalar degrees of freedom (and superpartners) near the intermediate 
scale. We also note vanishing corrections on the unification mass $(\U)$ due 
to intermediate-scale-threshold effects.\par
From analytic formulas, the threshold-corrections for mass scales are
\be\D \ln{\I\over \Z}&=&{(A'J_\D-AK_\D)\over (AB'-A'B)},\label{eq20}\\
\D \ln{\U\over \Z}&=&{(BK_\D-B'J_\D)\over (AB'-A'B)}.\label{eq21}\ee
We assume the extended survival hypothesis to operate with the consequence that
all scalar components of an $SO(10)$ representation which do not contribute to
spontaneous symmetry breaking are superheavy. Only lighter degrees of freedom 
are those $G_{2213}$-components in $126\op\ov{126}$ or $16\op\ov{16}$ which
contribute to spontaneous symmetry breaking at $\I$. Similarly the lightest
scalar components with masses near $\Z$ are up and down type doublets 
originating from $\un{10}\subset SO(10)$. The coloured triplets in $\un{10}$
have masses near the GUT scale. We compute threshold effects on $\I$ and $\U$ 
using two different methods which have been adopted in the current 
literature:\\
(\ref{seca}) Effective mass parameters and effective SUSY threshold have been
introduced by Carena, Pokorski and Wagner \cite{29} which have been also 
exploited in studying threshold effects in minimal SUSY GUTs \cite{22}. 
Similarly SUSY $SU(5)$ GUT-threshold effects have also been investigated by 
Langacker and Polonsky \cite{22} by introducing another set of effective mass 
parameters near the GUT scale. For the present analysis we utilize the same 
set of effective mass parameters at the SUSY scale as in ref. \cite{22} but 
use two new sets of effective mass parameters at $\I$ and $\U$. Although the 
effective mass parameters corresponding to SUSY threshold has been determined 
approximately using experimental measurements or well known estimations of the 
actual masses, such determinations for the effective mass parameters at higher 
thresholds has not been carried out due to lack of experimental data or 
adequate estimations on superheavy masses. In view of this we adopt the 
procedure similar to that outlined in \cite{22} and assume these effective 
mass parameters to be few times heavier or lighter than the corresponding mass 
scales.\\
(\ref{secb}) Without introducing effective mass parameters, threshold effects 
have been also computed conventionally by assigning specific and plausible 
values of masses to the superheavy scalar components in nonSUSY GUTs as well 
as SUSY theories \cite{30,31,32,33,34}. This method will be adopted below in a 
separate analysis. Following a result due to Shifman, masses used for 
estimation of threshold effects have been assumed to be bare masses as the 
wave function renormalization has been shown to get cancelled by two-loop 
effects \cite{35}.\par
In both these cases we find interesting solutions even when the masses are 
assigned their expected values and taken to be few times heavier or lighter 
than the corresponding scales.
\subsection{Threshold Effects with Effective Mass Parameters}\label{seca}
Including threshold corrections, we have investigated three models 
corresponding to $(n_{16}, n_{126})=$ (1, 0), (2, 0), (0, 1) with ${\un 
{45}}\op {\un {54}}$, for $SO(10)$ breaking and other three models 
corresponding to $(n_{16}, n_{126})=$ (1, 0), (2, 0), (0, 1) with ${\un 
{210}}$. The superheavy components in these models having masses near $\I$ and 
$\U$ are shown in Tables \ref{tab2} and \ref{tab3}, respectively. It is clear 
that threshold effects on $\U$ and $\I$ can be estimated once $M'_i (i 
=1Y, 2L, 3C)$ and $M''_i (i = 2L, 2R, BL, 3C)$ as defined in eqs. 
(\ref{eq22})-(\ref{eq24}) below are known. In any model the superheavy masses 
near any particular symmetry breaking scale can be parametrized in terms of 
the corresponding effective mass parameters \cite{22,29}. In the present model 
there are three such relations corresponding to the three symmetry breaking 
scales i.e., $\mu = M_{SUSY} = \Z$, $\mu=\I$ and $\mu=\U$,
\be \D^Z_i&=&\sum_{\a} {b^{\a}_i\over 2\pi}\ln{M_{\a}\over \Z}
={b_i\over 2\pi}\ln{M_i\over \Z},\nonumber\\
i&=&1Y,2L,3C;\, \mu=\Z;\label{eq22}\\
\D^I_i&=&\sum_{\a} {b'^{\a}_i\over 2\pi}\ln{M'_{\a}\over \I}
={b'_i\over 2\pi}\ln{M'_i\over \I},\nonumber\\ 
i&=&1Y,2L,3C;\, \mu=\I;\label{eq23}\\
\D^U_i&=&\sum_{\a} {b''^{\a}_i\over 2\pi}\ln{M''_{\a}\over \U}
={b''_i\over 2\pi}\ln{M''_i\over \U},\nonumber\\ 
i&=&2L,2R,BL,3C;\, \mu=\U;\label{eq24}\ee
where $\a$ refers to the actual $G_{213}$ submultiplet near $\mu = \Z$, $\I$ or
$G_{2213}$ submultiplet near $\mu = \U$ and $M_{\a}, M'_{\a}$ or $M''_{\a}$
refer to the actual component masses. The coefficients $b'_i=\sum b'^{(\a)}_i$ 
and $b''_i=\sum b''^{(\a)}_i$ have been defined in (\ref{eq22})-(\ref{eq24}) 
following refs. \cite{22,29}. The numbers $b^{\a}_i$ refer to the one-loop 
coefficients of the multiplet $\a$ under the gauge subgroup $U(1)_Y$, 
$SU(2)_L$, $SU(2)_R$, $SU(3)_C$, $U(1)_{B-L}$ etc. Using eqs.(\ref{eq9}) and 
(\ref{eq20})-(\ref{eq24}) we have obtained contributions to threshold effects 
on the two mass scales, $\D\ln{\I\over \Z}$ and $\D\ln{\U\over \Z}$, as shown in 
Table \ref{tab4} in terms of effective mass parameters $M'_i (i =1Y, 2L, 3C)$ 
and $M''_i (i = 2L, 2R, BL, 3C)$. The numerical entries in Table \ref{tab4} 
denote threshold effects at $\Z$ estimated using the effective mass parameters 
of ref. \cite{22}.\par
Using one-loop coefficients from Tables \ref{tab2}-\ref{tab3} and effective
mass parameters denoted as primes at the intermediate scale and as double
primes at the GUT scale, the analytic expressions for threshold corrections
are presented in Table \ref{tab4} where different models have been also
defined. A remarkable feature is that $\D \ln{\U\over \Z}$ has vanishing
corrections due to intermediate-scale threshold effects as the corresponding 
expressions contain no term involving any of the parameters like $M_{2L}'$, 
$M_{2R}'$, $M_{BL}'$ or $M_{3C}'$. Further, the effective mass parameters 
$M''_{2R}$ and $M''_{BL}$ have vanishing contributions to the threshold 
effects on the unification mass. Another notable feature is that corrections 
due to $M'_{2L}$ and $M'_{3C}$ are absent in $\D \ln{\I\over \Z}$. There is only 
a small correction due to $M'_{1Y}$.\par
The relation (\ref{eq22}) has been utilized in ref \cite{22} to compute only 
one set of values of $M_1$, $M_2$, $M_3$ in MSSM from the model predictions on
$M_{\a}$. But, since such predictions are also model dependent, several other 
assumed values of effective mass parameters have been utilized for computation.
At present no experimental or theoretical information is available on the 
actual values of superheavy masses around $\I$ and $\U$, although theoretically
it is natural to assume these masses to spread around the corresponding scales
by a factor bounded by ${1\over {10}}$ to 10. In the present case, in the absence
of actual values of component masses in the model, we make quite plausible 
and reasonable assumptions on $M'_i$ and $M''_i$ for computation. In our 
analysis the effective mass parameters $M'_i$ or $M''_i$ are taken to vary 
between ${1\over 5}-5$ times the relevant scale of symmetry breaking i.e., $\I$ 
or $\U$. Numerical solutions to different allowed values of mass scales 
corresponding to different choices of effective mass parameters are presented 
in Table \ref{tab5}.\par
Certain important features of these solutions are noteworthy. The minimal 
model-I with one set of $\un{210},\un{16}\op\un{\ov{16}}$, and $\un{10}$ 
permits intermediate scale in the interesting range of $10^3-10^{13}$ GeV for 
reasonable choices of effective mass parameters having string scale unification
$\U\simeq (5-6)\t 10^{17}$ GeV. Also the $SO(10)$ model with $\un{210},
\un{126}\op\un{\ov{126}}$, and $\un{10}$ allows $\I\simeq 5.3\t 10^{11}$ GeV 
with high unification mass close to the string scale $\U\simeq 5.6\t 10^{17}$ 
GeV. We also note that the intermediate scale solution with $\I\simeq 
10^{11}-10^{13}$ GeV is maintained irrespective of the fact whether a 
$\un{210}$ or a $\un{45}\op\un{54}$ or even a $\un{45}\op\un{210}$ are used 
for the GUT-scale symmetry breaking.\par 
Although solutions with intermediate scale $\I\simeq 10^{11}-10^{13}$ GeV are
also possible due to threshold effects with superheavy masses as shown in Sec.
\ref{secb}, a special and notable feature with effective mass parameters is
the possibility of low-mass right-handed gauge bosons corresponding to
$\I\simeq 1-10$ TeV in all the six models, minimal or nonminimal. These
solutions are also indicated in Table \ref{tab5}. Such low-mass right-handed
gauge bosons might be testified through experimentally detectable $V+A$ 
currents in future \cite{6}.
\subsection{Threshold Effects with Superheavy Masses}\label{secb}
In this subsection, instead of using effective mass parameters, we estimate 
threshold effects with reasonable choices on values of the masses of 
superheavy components of Higgs scalars and their superpartners in three models 
corresponding to $(n_{16}, n_{126})=$ (1, 0), (2, 0), (0, 1) with 
${\un {45}}\op {\un {54}}$, and three other models corresponding to 
$(n_{16}, n_{126})=$ (1, 0), (2, 0), (0, 1) with ${\un {210}}$.\par
The expression for threshold effect in terms of actual superheavy component 
masses $M'_{\a}$ which have values near $\I$ is given by eq. (\ref{eq23}),
\b \D^I_i=\sum_{\a} {b'^{\a}_i\over 2\pi}\ln{M'_{\a}\over \I},\,
i=1Y,2L,3C.\label{eq25}\e 
The superheavy components $\a$ contained in different models which have masses
near $\I$ are shown in Table \ref{tab2}.
Similarly, the threshold effect at $\mu = \U$ is given by eq. (\ref{eq24}),
\b \D^U_i=\sum_{\a} {b''^{\a}_i\over 2\pi}\ln{M''_{\a}\over \U},\, 
i=2L,2R,BL,3C.\label{eq26}\e
The superheavy components $\a$ contained in different models which have masses
near $\U$ are given in Table \ref{tab3}. While computing threshold effects, we
have assumed all the multiplets belonging to an $SO(10)$ representation `$H$' 
to have the same degenerate mass $M_H$. For example, all the superheavy 
components in $\un {45}$ given in Table \ref{tab3} near $\mu = \U$ have been 
subjected to the following degeneracy condition
$$M''(3,1,0,1) = M''(1,3,0,1) = M''(1,1,0,8) = M_{45}.$$
Similarly for $\un {210}$, $\un {54}$, $\un{16}\op\un{\ov {16}}$,
$\un{126}\op\un{\ov {126}}$ and $\un{10}$ 
\be &&M''(2,2,\pm{1\over 3},6)=M''(2,2,\pm{1\over 3},3) = ........ = M_{210},
\nonumber\\
&&M''(3,3,0,1)=M''(1,1,\pm{2\over 3},6) = ........ = M_{54},\nonumber\\
&&M''(2,1,-{1\over 2},1)=M''(2,1,{1\over 6},3) = ........ = M_{16},\nonumber\\
&&M''(1,3,-{1\over 3},\ov 6)=M''(1,3,{1\over 3},\ov 3) = ........ = M_{126},
\nonumber\\
&&M''(1,1,\pm{1\over 3},3)=M_{10}.\nonumber\ee
All the heavy masses near $\mu =\I$ are assumed to have the same mass $M'$. We 
have obtained contributions to threshold effects on the two mass scales, 
$\D\ln{\I\over \Z}$ and $\D\ln{\U\over \Z}$ as shown in Table \ref{tab4}, in terms 
of superheavy masses. The last term i.e., the numerical entries denote 
threshold contributions at $\mu = \Z$. From Table \ref{tab6}, it can be seen 
that the threshold contributions due to the superheavy masses $M_{16}$ or 
$M_{126}$ to $\D\ln{\U\over \Z}$ are absent for all the models and the threshold 
contribution due to the representation $\un {54}$ cancels out from models IV, 
V and VI. These cancellations are understood due to a theorem by Mohapatra 
\cite{36}. As before the masses are allowed to vary between ${1\over 5}$ and 5 
times the scale of relevant symmetry breaking i.e. $\I$ or $\U$. Numerical 
solutions to different allowed values of mass scales corresponding to 
different choices of superheavy masses are presented in Table \ref{tab7}. In 
Table \ref{tab7}, for models I, IV (II, V) we have intermediate scale 
$\I\simeq 10 ^{14}(10^{13})$ GeV with unification scale $\U\simeq 10 ^{16}$GeV
for reasonable choice of superheavy masses. In case of models III, VI the 
intermediate scale can be as low as $10^{10}$GeV with unification scale
$6.3\t 10^{15}$ GeV.
\section{Summary and Conclusion}\label{sec5}
While investigating the possibility of $G_{2213}$ intermediate gauge symmetry 
in SUSY $SO(10)$, we have considered a number of models. At the two-loop level,
we have noted that except for the nonminimal models with $3(\un{16}\op
{\un{\ov{16}}})$, $\un{210}$, $\un{10}$ \cite{10} and those of ref. 
\cite{13,14,16} the RGEs in the minimal models do not permit $G_{2213}$ 
intermediate breaking scale $\I$ to be substantially lower than $\U$ when both
gravitational and threshold effects are ignored. Including gravitational 
corrections in the minimal model, we observe that the prediction on 
unification mass remains unaffected by such Planck-scale-induced gravitational 
effects whereas the intermediate scale can be lowered by atmost two orders of 
magnitude through such corrections.\par
Including threshold corrections, we have considered various minimal and 
nonminimal models and obtained $\I\simeq 10^{10}-10^{13}$ GeV with high 
unification scale using plausible values of effective mass parameters but 
without using additional number of Higgs scalars at $\I$ (models I, III, IV 
and VI with $n_{126}=1$ or $n_{16}=1$). Our choices of effective mass 
parameters are similar to those used in earlier investigations \cite{22,29} 
and choices of superheavy masses near thresholds are similar to those used in 
earlier analyses \cite{30,31,32,33,34}. An important feature of analytic 
result is that the GUT scale is unaffected by the spreading of masses near the
intermediate scale although intermediate scale is itself changed significantly
by the superheavy masses near the GUT scale. Thus the proton lifetime 
predictions in the model for $p\to e+\pi^0$ mode are unchanged by 
gravitational or intermediate scale threshold corrections. Another important 
aspect of this analysis is that even if the spreading of masses near the two 
thresholds are only few times heavier or lighter than the corresponding 
scales, the models result in $\I\simeq 10^{11}-10^{13}$ GeV either with 
effective mass parameters or with superheavy masses. We further observe that 
low-mass right-handed gauge bosons in the range $1-10$ TeV are permitted in 
the model only when threshold effects are computed with effective mass 
parameters. All relevant superheavy masses contributing to threshold effects
have been assumed to be bare masses as their wave function renormalisation has
been shown to be cancelled out by two-loop effects \cite{35}.\par
It may be noted that while the use of $\un{126}\op\un{\ov{126}}$ permits the 
implementation of conventional see-saw mechanism for neutrino masses with 
$R$-parity conservation, it is possible to use a generalized mechanism 
\cite{37,38} with the choice $\un{16}\op\un{\ov{16}}$, such that one can get a 
see-saw like formula for light neutrino masses. In the latter case $R$-parity 
is violated and one needs to impose additional discrete symmetries to maintain 
the stability of the proton. We thus conclude that, $G_{2213} (g_{2L}\neq 
g_{2R})$ with minimal choice of Higgs scalars is allowed as an intermediate 
gauge symmetry in SUSY $SO(10)$ model. The right-handed Majorana neutrinos 
associated with the intermediate scales obtained in this analysis are 
compatible with observed indications for light Majorana neutrino masses 
through see-saw mechanism \cite{5,8,9}.\par
Recently interesting investigations have been made in SUSY left-right gauge
models while embedding $G_{2213}$ in SUSY $SO(10)$ as an intermediate gauge 
group in the presence of higher dimensional operators \cite{39}. It would be 
more interesting to study the impact of threshold effects in such models where 
certain light degrees of freedom are allowed naturally and $R$-parity 
conservation is guaranteed.
\begin{acknowledgement} The works of M.K.P and C.R.D. are supported by the 
project No. SP/S2/K-30/98 of Department of Science and Technology, Govt. of 
India.\end{acknowledgement}
\begin{table*}
\caption{Mass scales and coefficients for different SUSY $SO(10)$ models 
including one-loop and two-loop contributions. Also shown are the values of 
$\I$ including the 5-dim. operator effect while $\U$ remains unaffected.}
\label{tab1}
\begin{tabular}{cccccccccccc}\hline\noalign{\smallskip}
$n_{16}$&$n_{126}$&$A$&$B$&$A'$&$B'$&One-loop&One-loop&Two-loop&Two-loop&
5-dim. operator&$\I$\\
&&&&&&$\I$ (GeV)&$\U$ (GeV)&$\I$ (GeV)&$\U$ (GeV)&$\eta$&(GeV)\\  
\noalign{\smallskip}
\hline\noalign{\smallskip}
1&0&${16\over 3}$&4&16&4&$10^{15.96}$&$10^{16.5}$&$10^{16.9}$&$10^{16.11}$&8&
$10^{15.9}$\\\noalign{\smallskip}
$0$&$1$&${40\over 3}$&$-4$&$24$&$-4$&$10^{17.00}$&$10^{16.5}$&$10^{15.2}$&
$10^{16.11}$&$-8$&$10^{14.3}$\\\noalign{\smallskip}
$2$&$0$&${22\over 3}$&$2$&$18$&$2$&$10^{15.44}$&$10^{16.5}$&$10^{17.6}$&
$10^{16.12}$&$8$&$10^{15.8}$\\\noalign{\smallskip}
$0$&$2$&${70\over 3}$&$-14$&$34$&$-14$&$10^{16.68}$&$10^{16.5}$&$10^{15.92}$&
$10^{16.11}$&$-8$&$10^{15.6}$\\\noalign{\smallskip}
$1$&$1$&${46\over 3}$&$-6$&$26$&$-6$&$10^{16.83}$&$10^{16.5}$&$10^{15.5}$&
$10^{16.09}$&$-8$&$10^{14.9}$\\\noalign{\smallskip}
$2$&$1$&${52\over 3}$&$-8$&$28$&$-8$&$10^{16.74}$&$10^{16.5}$&$10^{15.69}$&
$10^{16.12}$&$-8$&$10^{15.2}$\\\noalign{\smallskip}
$1$&$2$&${76\over 3}$&$-16$&$36$&$-16$&$10^{16.61}$&$10^{16.5}$&$10^{15.9}$&
$10^{16.12}$&$-8$&$10^{15.6}$\\\noalign{\smallskip}
$0$&$3$&${100\over 3}$&$-24$&$44$&$-24$&$10^{16.57}$&$10^{16.5}$&$10^{16.9}$&
$10^{15.90}$&$-8$&$10^{15.8}$\\\noalign{\smallskip}
$3$&$0$&${28\over 3}$&$0$&$20$&$0$&---&$10^{16.5}$&---&---&---&---\\
\noalign{\smallskip}\hline\end{tabular}\end{table*}
\begin{table}
\caption{The heavy Higgs content of the $SO(10)$ model with $G_{2213}$ 
intermediate symmetry. The $G_{213}$ submultiplets acquire masses close to 
$\I$ when $G_{2213}$ is broken.}\label{tab2}
\begin{tabular}{ccc}\hline\noalign{\smallskip}
$SO(10)$ representation&$G_{213}$ multiplet&$b'_{2L}, b'_{1Y}, b'_{3C}$\\
\noalign{\smallskip}\hline\noalign{\smallskip}
$\un{16}$&$(1, 0, 1)$&$(0, 0, 0)$\\\noalign{\smallskip}
$\un{\ov{16}}$&$(1, -1, 1)$&$(0, {3\over 5}, 0)$\\ \\
$\un{16}$&$(1, 0, 1)$&$(0, 0, 0)$\\\noalign{\smallskip}
$\un{\ov{16}}$&$(1, -1, 1)$&$(0, {3\over 5}, 0)$\\\noalign{\smallskip}
$\un{16}\left[\un{\ov{16}}\right]$&$(1, 0, 1)$&$(0, 0, 0)$\\
\noalign{\smallskip}
&$(1, 1, 1)[(1, -1, 1)]$&$(0, {3\over 5}, 0)$\\ \\
$\un{126}$&$(1, 2, 1)$&$(0, {12\over 5}, 0)$\\\noalign{\smallskip}
&$(1, 0, 1)$&$(0, 0, 0)$\\\noalign{\smallskip}
$\un{\ov{126}}$&$(1, -1, 1)$&$(0, {3\over 5}, 0)$\\\noalign{\smallskip}
&$(1, -2, 1)$&$(0, {12\over 5}, 0)$\\\noalign{\smallskip}\hline
\end{tabular}\end{table}
\begin{table*}
\caption{Same as Table II, but here the $G_{2213}$ submultiplets acquire 
masses close to $\U$ when $SO(10)$ is broken.}\label{tab3}
\begin{tabular}{ccc}\hline\noalign{\smallskip}
$SO(10)$ representation&$G_{2213}$ submultiplet&$b''_{2L}, b''_{2R}, 
b''_{BL}, b''_{3C}$\\\noalign{\smallskip}
\hline\noalign{\smallskip}
$\un{210}$&$(2, 2, \pm{1\over 3}, 6)$&$(6, 6, 4, 10)$\\\noalign{\smallskip}
&$(2, 2, \pm{1\over 3}, 3)$&$(3, 3, 2, 2)$\\\noalign{\smallskip}
&$(2, 2, \pm 1, 1)$&$(1, 1, 6, 0)$\\\noalign{\smallskip}
&$(3, 1, \pm{2\over 3}, 3)$&$(6, 0, 6, {3\over 2})$\\\noalign{\smallskip}
&$(1, 3, \pm{2\over 3}, 3)$&$(0, 6, 6, {3\over 2})$\\  \noalign{\smallskip}
&$(3, 1, 0, 8)$&$(16, 0, 0, 9)$\\\noalign{\smallskip}
&$(1, 3, 0, 8)$&$(0, 16, 0, 9)$\\\noalign{\smallskip}
&$(3, 1, 0, 1)$&$(2, 0, 0, 0)$\\\noalign{\smallskip}
&$(1, 3, 0, 1)$&$(0, 2, 0, 0)$\\\noalign{\smallskip}
&$(1, 1, 0, 8)$&$(0, 0, 0, 3)$\\ \\
$\un{45}$&$(3, 1, 0, 1)$&$(2, 0, 0, 0)$\\\noalign{\smallskip}
&$(1, 3, 0, 1)$&$(0, 2, 0, 0)$\\\noalign{\smallskip}
&$(1, 1, 0, 8)$&$(0, 0, 0, 3)$\\ \\
$\un{54}$&$(3, 3, 0, 1)$&$(6, 6, 0, 0)$\\\noalign{\smallskip}
&$(1, 1, \pm{2\over 3}, 6)$&$(0, 0, 4, {5\over 2})$\\\noalign{\smallskip}
&$(1, 1, 0, 8)$&$(0, 0, 0, 3)$\\ \noalign{\smallskip}
&$(2, 2, \pm{1\over 3}, 3)$&$(3, 3, 2, 2)$\\ \\
$\un{16}\left[\un{\ov{16}}\right]$&$(2, 1, -{1\over 2}, 1)[(2, 1, {1\over 2}, 1)]$&
$({1\over 2}, 0, {3\over 4}, 0)$\\\noalign{\smallskip}
&$(2, 1, {1\over 6}, 3)[(2, 1, -{1\over 6}, \ov 3)]$&$({3\over 2}, 0, {1\over 4}, 1)$\\
\noalign{\smallskip}    
&$(1, 2, -{1\over 6}, 3)[(1, 2, {1\over 6}, \ov 3)]$&$(0, {3\over 2}, {1\over 4}, 1)$\\ \\
$\un{126}\left[\un{\ov{126}}\right]$&$(1, 3, -{1\over 3}, \ov 6)[(1, 3, {1\over 3}, 
6)]$&$(0, 12, 3, {15\over 2})$\\\noalign{\smallskip}
&$(1, 3, {1\over 3}, \ov 3)[(1, 3, -{1\over 3}, 3)]$&$(0, 6, {3\over 2}, {3\over 2})$\\
\noalign{\smallskip}    
&$(3, 1, {1\over 3}, 6)[(3, 1, -{1\over 3}, \ov 6)]$&$(12, 0, 3, {15\over 2})$\\
\noalign{\smallskip}    
&$(3, 1, -{1\over 3}, 3)[(3, 1, {1\over 3}, \ov 3)]$&$(6, 0, {3\over 2}, {3\over 2})$\\
\noalign{\smallskip}    
&$(3, 1, -1, 1)[(3, 1, 1, 1)]$&$(2, 0, {9\over 2}, 0)$\\\noalign{\smallskip}    
&$(2, 2, {2\over 3}, 3)[(2, 2, 1, -{2\over 3}, \ov 3)]$&$(3, 3, 8, 2)$\\
\noalign{\smallskip}    
&$(2, 2, -{2\over 3}, \ov 3)[(2, 2, {2\over 3}, 3)]$&$(3, 3, 8, 2)$\\
\noalign{\smallskip}    
&$(2, 2, 0, 8)[(2, 2, 0, 8)]$&$(8, 8, 0, 12)$\\\noalign{\smallskip}    
&$(2, 2, 0, 1)[(2, 2, 0, 1)]$&$(1, 1, 0, 0)$\\\noalign{\smallskip}
&$(1, 1, -{1\over 3}, 3)[(1, 1, {1\over 3}, \ov 3)]$&$(0, 0, {1\over 2}, {1\over 2})$\\
\noalign{\smallskip}    
&$(1, 1, {1\over 3}, \ov 3)[(1, 1, -{1\over 3}, 3)]$&$(0, 0, {1\over 2}, {1\over 2})$\\ \\
$\un{10}$&$(1, 1, \pm{1\over 3},3)$&$(0, 0, {1\over 2}, {1\over 2})$\\ 
\noalign{\smallskip}\hline\end{tabular}\end{table*}
\begin{table*}
\caption{Threshold effects on mass scales in different SUSY $SO(10)$ models 
using effective mass parameters.}\label{tab4}
\begin{tabular}{cccc}\hline\noalign{\smallskip}
MODEL&Representation content&$\D \ln{\I\over \Z}$&$\D \ln{\U\over \Z}$ \\
\noalign{\smallskip}\hline\noalign{\smallskip}
I&$\un{210}, \un{16}\op\un{\ov {16}}, \un{10}$&${53\over 4}\ln{M''_{2R}\over \U}
+{103\over 12}\ln{M''_{BL}\over \U}$&${27\over 2}\ln{M''_{2L}\over \U}
-14\ln{M''_{3C}\over \U}$\\\noalign{\smallskip}
&&$-{81\over 2}\ln{M''_{2L}\over \U}+{56\over 3}\ln{M''_{3C}\over \U}$&$+0.117$\\
\noalign{\smallskip}
&&$-1.33$&\\ \\
II&$\un{210}, 2(\un{16}\op\un{\ov {16}}), \un{10}$&$28\ln{M''_{2R}\over \U}
+18\ln{M''_{BL}\over \U}$&${29\over 2}\ln{M''_{2L}\over \U}-15\ln{M''_{3C}\over \U}$\\
\noalign{\smallskip}
&&$-{203\over 2}\ln{M''_{2L}\over \U}+55\ln{M''_{3C}\over\U}$&$+0.122$\\
\noalign{\smallskip}
&&$+{3\over 2}\ln{M'_{1Y}\over \I}-2.78$&\\ \\
III&$\un{210}, \un{126}\op\un{\ov {126}}, \un{10}$&$-29\ln{M''_{2R}\over\U}
-{55\over 3}\ln{M''_{BL}\over\U}$&$30\ln{M''_{2L}\over\U}-{121\over 4}\ln{M''_{3C}\over \U}$\\
\noalign{\smallskip}
&&$+150\ln{M''_{2L}\over\U}-{305\over 3}\ln{M''_{3C}\over\U}$&$+0.105$\\
\noalign{\smallskip}
&&$-{9\over 4}\ln{M'_{1Y}\over\I}-1.56$&\\ \\
IV&$\un{45},\un{54}, \un{16}\op\un{\ov {16}}, \un{10}$&
${5\over 4}\ln{M''_{2R}\over \U}+{7\over 12}\ln{M''_{BL}\over \U}$&
${3\over 2}\ln{M''_{2L}\over \U}-2\ln{M''_{3C}\over \U}$\\\noalign{\smallskip}
&&$-{9\over 2}\ln{M''_{2L}\over \U}+{8\over 3}\ln{M''_{3C}\over \U}$&$+0.117$\\
\noalign{\smallskip}
&&$-1.33$&\\ \\
V&$\un{45}, \un{54}, 2(\un{16}\op\un{\ov {16}}), \un{10}$&$4\ln{M''_{2R}\over \U}
+2\ln{M''_{BL}\over \U}$&${5\over 2}\ln{M''_{2L}\over \U}-3\ln{M''_{3C}\over \U}$\\
\noalign{\smallskip}
&&$-{35\over 2}\ln{M''_{2L}\over \U}+11\ln{M''_{3C}\over\U}$&$+0.122$\\
\noalign{\smallskip}
&&$+{3\over 2}\ln{M'_{1Y}\over \I}-2.78$&\\ \\
VI&$\un{45},\un{54}, \un{126}\op\un{\ov {126}}, \un{10}$&$-17\ln{M''_{2R}\over\U}
-{31\over 3}\ln{M''_{BL}\over\U}$&$18\ln{M''_{2L}\over\U}-{37\over 2}\ln{M''_{3C}\over \U}$\\
\noalign{\smallskip}
&&$+90\ln{M''_{2L}\over\U}-{185\over 3}\ln{M''_{3C}\over\U}$&$+0.105$\\
\noalign{\smallskip}
&&$-{9\over 4}\ln{M'_{1Y}\over\I}-1.56$&\\
\noalign{\smallskip}\hline\end{tabular}\end{table*}
\begin{table*}
\caption{Predictions on mass scales $\I$ and $\U$ including threshold effect 
with effective mass parameters.}\label{tab5}
\begin{tabular}{ccccccccc}\hline\noalign{\smallskip}
Model&Two-loop (GeV)&$M'_{1Y}$&$M''_{2L}$&$M''_{2R}$&$M''_{BL}$&$M''_{3C}$&
$\I$(GeV)&$\U$(GeV)\\\noalign{\smallskip}\hline\noalign{\smallskip}
I&$\I=10^{16.9}$&-&$4\U$&$3\U$&$3\U$&$2.95\U$&$1.36\t10^{11}$&
$5.2\t10^{17}$\\\noalign{\smallskip}
&$\U=10^{16.11}$&-&$5\U$&$3.5\U$&$3.5\U$&$\U$&$2\t10^{11}$&
$1.5\t10^{17}$\\\noalign{\smallskip}
&&-&$3\U$&$\U$&$\U$&$2.2\U$&$1.6\t10^{3}$&$6.5\t10^{17}$\\ \\
II&$\I=10^{17.6}$&$\I$&$3\U$&$4\U$&$2\U$&$2.5\U$&$1.0\t10^{11}$&
$1.3\t10^{17}$\\\noalign{\smallskip}
&$\U=10^{16.12}$&$\I$&$3\U$&$3.3\U$&$3.3\U$&$2.27\U$&$2.78\t10^{11}$&
$5.58\t10^{17}$\\\noalign{\smallskip}
&&${1\over 3}\I$&$4\U$&$5\U$&$2\U$&$3\U$&$1.3\t10^{3}$&$5.55\t10^{17}$\\ \\
III&$\I=10^{15.2}$&$\I$&$\U$&$\U$&$2\U$&$\U$&$1.11\t10^{9}$&
$1.46\t10^{16}$\\\noalign{\smallskip}
&$\U=10^{16.11}$&$3.5\I$&$2.27\U$&$3.5\U$&$3.5\U$&$2\U$&$5.3\t10^{11}$&
$5.5\t10^{17}$\\\noalign{\smallskip}
&&$\I$&$\U$&$\U$&$5\U$&$\U$&$1.2\t10^{3}$&$1.46\t10^{16}$\\ \\
IV&$\I=10^{16.9}$&-&$2\U$&$\U$&$\U$&$\U$&$1.84\t10^{12}$&$3.34\t10^{17}$\\
\noalign{\smallskip} 
&$\U=10^{16.11}$&-&$2\U$&$\U$&$\U$&${1\over {1.5}}\U$&$1.23\t10^{11}$&
$2.53\t10^{18}$\\\noalign{\smallskip}
&&-&$3\U$&${1\over 5}\U$&${1\over 5}\U$&${1\over 2}\U$&$1.2\t10^{3}$&
$6.6\t10^{19}$\\ \\
V&$\I=10^{17.6}$&$\I$&$\U$&${1\over {1.5}}\U$&$5\U$&${1\over 2}\U$&
$1.77\t10^{12}$&$9.94\t10^{17}$\\\noalign{\smallskip}
&$\U=10^{16.12}$&$\I$&$\U$&${1\over 1.2}\U$&${1\over 5}\U$&$\U$&
$2.84\t10^{11}$&$1.48\t10^{16}$\\\noalign{\smallskip}
&&${1\over 5}\I$&$2\U$&$\U$&$\U$&$\U$&$6.3\t10^{3}$&$6.7\t10^{17}$\\ \\
VI&$\I=10^{15.2}$&$5\I$&${1\over 2}\U$&$\U$&${1\over 5}\U$&$\U$&$1.0\t10^{14}$&
$6.75\t10^{17}$\\\noalign{\smallskip}
&$\U=10^{16.11}$&$\I$&$\U$&${1\over 1.2}\U$&$3\U$&$\U$&$4\t10^{11}$&
$1.46\t10^{16}$\\\noalign{\smallskip}
&&$5\I$&$\U$&${1\over 1.1}\U$&$5\U$&$1.1\U$&$3.7\t10^{3}$&$1.87\t10^{15}$\\
\noalign{\smallskip}\hline\end{tabular}\end{table*}
\begin{table*}
\caption{Threshold effects on mass scales in different SUSY $SO(10)$ models
using superheavy masses.}\label{tab6}
\begin{tabular}{cccc}\hline\noalign{\smallskip}
MODEL&Representation content&$\D \ln{\I\over \Z}$&$\D \ln{\U\over \Z}$ \\
\noalign{\smallskip}\hline\noalign{\smallskip}
I&$\un{210}, \un{16}\op\un{\ov {16}}, \un{10}$&$-{1\over 2}\ln{M_{16}\over \U}
+{1\over 2}\ln{M_{10}\over \U}$&$-{1\over 4}\ln{M_{210}\over \U}-{1\over 4}\ln{M_{10}\over \U}$\\
\noalign{\smallskip}
&&$-1.33$&$+0.117$\\ \\
II&$\un{210}, 2(\un{16}\op\un{\ov {16}}), \un{10}$&${1\over 4}\ln{M_{210}\over \U}
-2\ln{M_{16}\over \U}$&$-{1\over 4}\ln{M_{210}\over \U}-{1\over 4}\ln{M_{10}\over \U}$\\
\noalign{\smallskip}
&&$+{5\over 4}\ln{M_{10}\over \U}+\ln{M'\over \I}$&$+0.122$\\\noalign{\smallskip}
&&$-2.78$&\\ \\
III&$\un{210}, \un{126}\op\un{\ov {126}}, \un{10}$&$-{1\over 2}\ln{M_{210}\over \U}
+{5\over 2}\ln{M_{126}\over \U}$&$-{1\over 4}\ln{M_{210}\over \U}
-{1\over 4}\ln{M_{10}\over \U}$\\\noalign{\smallskip}
&&$-\ln{M_{10}\over \U}-2\ln{M'\over \I}$&$+0.105$\\\noalign{\smallskip}
&&$-1.56$&\\ \\
IV&$\un{45},\un{54}, \un{16}\op\un{\ov {16}}, \un{10}$&$
-{1\over 2}\ln{M_{16}\over \U}+{1\over 2}\ln{M_{10}\over \U}$&$-{1\over 4}\ln{M_{45}\over \U}
-{1\over 4}\ln{M_{10}\over \U}$\\\noalign{\smallskip}
&&$-1.33$&$+0.117$\\ \\
V&$\un{45}, \un{54}, 2(\un{16}\op\un{\ov {16}}), \un{10}$&
${1\over 4}\ln{M_{45}\over \U}-2\ln{M_{16}\over \U}$&$-{1\over 4}\ln{M_{45}\over \U}
-{1\over 4}\ln{M_{10}\over \U}$\\\noalign{\smallskip}
&&$+{5\over 4}\ln{M_{10}\over \U}+\ln{M'\over \I}$&$+0.122$\\\noalign{\smallskip}
&&$-2.78$&\\ \\
VI&$\un{45},\un{54}, \un{126}\op\un{\ov {126}}, \un{10}$&
$-{1\over 2}\ln{M_{45}\over\U}-{5\over 2}\ln{M_{126}\over\U}$&
$-{1\over 4}\ln{M_{45}\over\U}-{1\over 4}\ln{M_{10}\over \U}$\\\noalign{\smallskip}
&&$-\ln{M_{10}\over \U}-2\ln{M'\over \I}$&$+0.105$\\\noalign{\smallskip}
&&$-1.56$&\\\noalign{\smallskip}\hline\end{tabular}\end{table*}
\begin{table*}
\caption{Predictions on mass scales $\I$ and $\U$ including threshold effect 
with superheavy masses.}\label{tab7}
\begin{tabular}{ccccccccc}\hline\noalign{\smallskip}
Model&Two-loop(GeV)&$M'$&$M_{210}${\rm or}$M_{45}$&$M_{16}$&$M_{126}$&
$M_{10}$&$\I$(GeV)&$\U$(GeV)\\\noalign{\smallskip}\hline\noalign{\smallskip}
I,IV&$\I=10^{16.9}$&-&${1\over 5}\U$&${1\over 5}\U$&-&${1\over 5}\U$&
$2.1\t10^{16}$&$3.2\t10^{16}$\\\noalign{\smallskip}
&$\U=10^{16.11}$&-&${1\over 5}\U$&$5\U$&-&${1\over 5}\U$&$9.7\t10^{14}$&
$3.2\t10^{16}$\\ \\
II,V&$\I=10^{17.6}$&${1\over 5}\I$&${1\over 5}\U$&${1\over 5}\U$&-&
${1\over 5}\U$&$1.1\t10^{16}$&$3.3\t10^{16}$\\\noalign{\smallskip}
&$\U=10^{16.12}$&${1\over 5}\I$&${1\over 5}\U$&$5\U$&-&${1\over 5}\U$&
$1.7\t10^{13}$&$3.3\t10^{16}$\\ \\
III,VI&$\I=10^{15.2}$&${1\over 5}\I$&$5\U$&-&${1\over 5}\U$&$5\U$&
$2.3\t10^{10}$&$6.3\t10^{15}$\\\noalign{\smallskip}
&$\U=10^{16.11}$&${1\over 5}\I$&$5\U$&-&${1\over 5}\U$&$\U$&
$1.1\t10^{11}$&$9.5\t10^{15}$\\\noalign{\smallskip}\hline\end{tabular}
\end{table*}
\end{document}